# Evolutionary theory of convective organization

Brian Mapes, University of Miami

# Abstract

The conceptual landscape of convection has two simple gateways: optimal function and random form. Optimal convection adjusts toward a univariate ideal called neutrality. Convection's form involves elements (parcels, bubbles, drafts) whose most parsimonious assumption is random. Between these gates lies a wilderness of realizable flow configurations. The only simple principle is natural selection by "fitness", a scalar whose gradient is a local direction in an abstract *configuration space*. Random or high-entropy patterns occupy most of configuration space and occur spontaneously. With time, convection can discover less facile but more efficient ("organized") configurations, by sequential selection. Here two data exercises explore that self-organization process, in shallow and deep moist convection. For shallow convection, causal network postulates are explored in a large set of cyclic-domain large-eddy simulations (LES; the "Cloud Botany" set). When an evolutionary pathway (mainly layer deepening in these simulations) leads to precipitation, mesoscale patterns blossom rapidly. For deep convection, expanding rings of conditional cell probability around prior cells are estimated from satellite imagery over South America and the South Pacific. In a Monte Carlo model iterating such a conditional probability kernel, hundreds of hourly cells take days to discover a self-sustaining "squall" configuration the kernel affords. Larger-scale implications include overshoots (redefinition of neutrality) and tens-of-hours timescales to both adjustment and noise (indeterminacy). If functional organization can be inferred from horizontal patterns, the abundance of horizontal texture information in satellite cloud imagery could find quantitative value.

# 1. Introduction and motivation

Buoyancy is a powerful, precise, intimate density selection force acting on all scales simultaneously. In highly selected phenomena, randomness is a poor but convenient structural approximation, as a starting point. Another definite but limited concept is *optimum*, an asymptotic value of some one-dimensional scalar as in notions of equilibrium or neutrality. Complicated reality plays out in a vast space of possibility bounded by these simple ideals. In economics the space between these "gates" is depicted memorably as "the wilderness of bounded rationality" in Fig. 8 of Farmer (2025). Ecology has a similar concept of nonunique, historically contingent complex situations that are characterized by a propensity for some information measure to increase with time, in "successions" after disruptions (Ulanowicz 1997, 2009). Lotka's 1922 "third law" of thermodynamics formalizes Darwin's dangerous idea (Dennett 1996), painted with enormous generality by Wong et al. (2023) as a Law of Increasing Functional Information. A thorough review from a physical science perspective is Goldenfeld and Woese (2011).

When convective motions begin anew in a uniformly destabilized fluid, their succession starts from the least-complex (highest entropy; most probable) configurations, often as independently developing cells with a size roughly the depth of the convecting layer. As interactions spread, small biases in conditional probability can iteratively multiply in a cascading self-organization process. Meanwhile, broader mesoscale motions respond more slowly but directly to gravity's pull, warping and modulating the small aspect ratio cells. Those two scales are not separate: convecting mesoscale flows could be viewed as a *coalitional strategy* in game theory terms, out-competing (by conjecture) both isolated cells and laminar mesoscales alone. With time, the game survival selects flow configurations based on some combination of energy efficiency (leading to vigor) and structural robustness (fostering persistence). Reproduction is another form of configuration survival and can occur visibly (e.g. as discrete propagation in squall lines) or more covertly, for instance through resonances with clear-air internal waves (Balaji and Clark 1988, Lane and Zhang 2011, Stephan et al. 2021).

Defining a space of convective flow "configurations" is not straightforward. Descriptive categorizations of commonly observed "storms" undoubtedly hold clues to discovering successful functional configurations. But convection as a process always involves closed circulations, not just easily observed cloud patterns. As a first step, an evolutionary theory of self-organization needs to define or discover a useful abstract space of configurations. It must not be infinitely large, and must have pathways of "adjacent possible" (Steel et al. 2020) stepping-stones for the evolutionary process.

Gradual development of multi-scale self-organization is commonly observed in plan-view imagery and simulations of cloudy convection, both shallow and deep. Detailed descriptions and budgets of the *how* of dynamical interactions (e.g. Janssens et al. 2022, Vieweg 2024) do not necessarily exhaust systemic questions of *why*. Relating convection's function (the why) to observable forms remains a grand challenge (Mapes 2021), a fundamentally scientific problem but with practical implications (as emphasized in Arakawa's venerable 2004 review). Beyond numerical model parameterizations, practical issues include the dependence of precipitation statistics on probability kernels (e.g. Ahmed and Neelin 2019) and radiative climate impacts of shallow clouds (Xu et al. 2023, Alinaghi et al. 2024, Janssens et al. 2025). These pattern-evolution effects can show themselves far sooner but more subtly than the days-to-weeks total collapses emphasized in the literature of "self-aggregation" (Wing et al. 2017, Muller et al. 2022).

The *why* of convection is enforced by the vertical buoyancy force b in the Boussinesq or anelastic equations. Lowering the center of gravity of the atmosphere (measured by some "available" potential energy APE) is convection's reason for existing, its mission. That APE is converted to kinetic energy at a work rate measured by its spatial correlation with vertical velocity, [b'w']. For dry convection (where b is a conserved variable) b'w' is a flux, but this word is a misnomer for moist convection. The APE-reduction mission, driven by gravity, is subject to strong constraints: the fluid laws for mass continuity (enforced by pressure), and inertia in Newton's law. Moist convection also hinges on the intimate saturation contingency of b's dependence on (conserved) specific humidity q. No unique measure of "available" potential energy exists (Randall 19xx), limiting our ability to close a game-theory formulation of an energy budget in any unique or even temporally local manner. When is q upwind of a saturation event to be counted as "potential" or "available"?

For this reason, dry convection is a useful starting point. The term "flow configurations" could be expressed as sets of weights in an expansion on an orthogonal basis set, for instance Fourier modes. Many spectral studies emphasize the power spectrum, but the phase spectrum is surely more crucial to fitness, one we have no way of decoding. I return to this subject in section 4. Whatever the patterns that win, the bulk signature of evolution should be that convection's energy efficiency (in some sense) increases with time, as unlikely but more efficient configurations are discovered (or discover themselves). Attempts below (section 2) to seek such temporal efficiency trends in shallow cloud simulations are thwarted by complications of moisture and layer-depth growth, so a retreat to dry turbulence should be attampted.

Measures of cloud-field *form* (morphometrics) are often used as an observation-driven "basis" for configuration space, but these are not orthogonal nor complete. Arbitrary measures have proliferated precisely because of a lack of clear relationship to function (energy "mission" efficiency or fitness). Intercorrelations between organization morphometrics are far from diagonal (Table 1 in Brune et al. 2021; Sakaeda and Torri 2021; supplement of Shamekh et al 2023), so there is clearly more than one aspect. Codes for computing all metrics are compiled in Janssens et al. 2021, which usefully distills "four interpretable dimensions" to their space: Characteristic length, void size, directional alignment, and cloud top height variance. The colorful names settled on by Rash et al. (2018) – sugar, gravel, flowers, fish – also suggest that pattern space has about four natural dimensions. But these horizontal patterns are so entangled with convecting layer depth and the onset of precipitation that shallow moist convection may not be satisfyingly amenable to evolutionary analysis (as concluded in section 2).

Precipitating deep convection (section 3) may be conceptually simpler in that sense, ripe for horizontal pattern analysis. Simulation experiments in Tsai and Mapes (2025) are perhaps our best example of results that can only be understood in evolutionary terms: *hardship breeds hardiness, gradually* under mechanism denial in half a periodic domain with uniform forcing. This paper presents another example of evolutionary analysis of deep convection patterns, iterating 2D conditional probability kernels estimated from satellite imagery over South America. Configuration space enumerations (whose inverse is related to probability) can be glimpsed through such efforts, although more systematic work is needed.

Section 4 will return to these larger themes before conclusions in section 5.

# 2. Shallow cloud pattern evolution: Cloud Botany

Shallow cloud patterns in long-convecting airmasses over subtropical oceans are extremely diverse. They evolve for tens of hours or days on downwind trajectories from initially stable stratocumulus decks (Wood 2012) or cold-air outbreaks (Murray-Watson et al. 2023). The meaningful identity of boundary-layer cloud trajectories for at least 2 days has been well confirmed (e.g. Albrecht et al. 2019). Thanks to convolutional neural nets, multiscale pattern information in imagery tiles or patches can now be accessed agnostically, rather than by arbitrary morphometric algorithms of convenience. Such pattern information can then be interpreted, whether through visual resemblances to animals or plants or commodities in supervised mimicry (Rasp et all. 2020) or more agnostically in unsupervised classification (Denby 2020, 2023) machine-learning exercises. More powerful is learning how pattern relates to physical impacts (Shamekh et al. 2023, Alinaghi et al. 2024, McCoy et al. 2025), or at least to a meteorological situation indicator like airmass age (section 5.2 of Schulz et al. 2021).

Self-organizing mesoscales up to 160km in size emerge in large-eddy simulations (LES, Lamaakel et al. 2023), taking more than a day. The recent Cloud Botany simulation ensemble (Jansson et al. 2023) presents a new opportunity to explore scales almost that large in an open Web-hosted LES dataset of 103 three-night-long LES simulations. This article's open code notebooks were developed from that team's excellent open-source templates and examples. Each simulation has a 153.6 km square cyclic domain at 100m horizontal resolution, beginning from a uniform initial state over a uniform water surface. All pattern development is spontaneous, conditional only on prior patterns. Forcing and initial condition parameter sweeps were designed from an analysis of western Atlantic subtropical observations, yielding diverse outcomes representative of various flow regimes. Simulations span 3 nights with 2 stabilizing days of fully interactive radiation. Winds undergo inertial oscillations and adjustment as surface friction intrudes into initially geostrophic wind profiles at a subtropical latitude, sometimes producing shears that act to align clouds in rows.

## a. Energy and pattern feedbacks in a causal network

To motivate the work below, some conceptual framing is helpful. Figure 1 shows a traditional causal network of convection, viewed as a process to be parameterized for larger horizontal scale dynamics among atmospheric "columns" the width of LES domain averages (indicated by the overbar). In the actual cyclic LES, of course, those larger scales are mathematically absent. The basic paradigm of Fig. 1 can be considered even for dry (unsaturated) convection of a cooled fluid over a warm surface, although "waves" exist only if that convection exists under a stratified layer like moist convection generates.

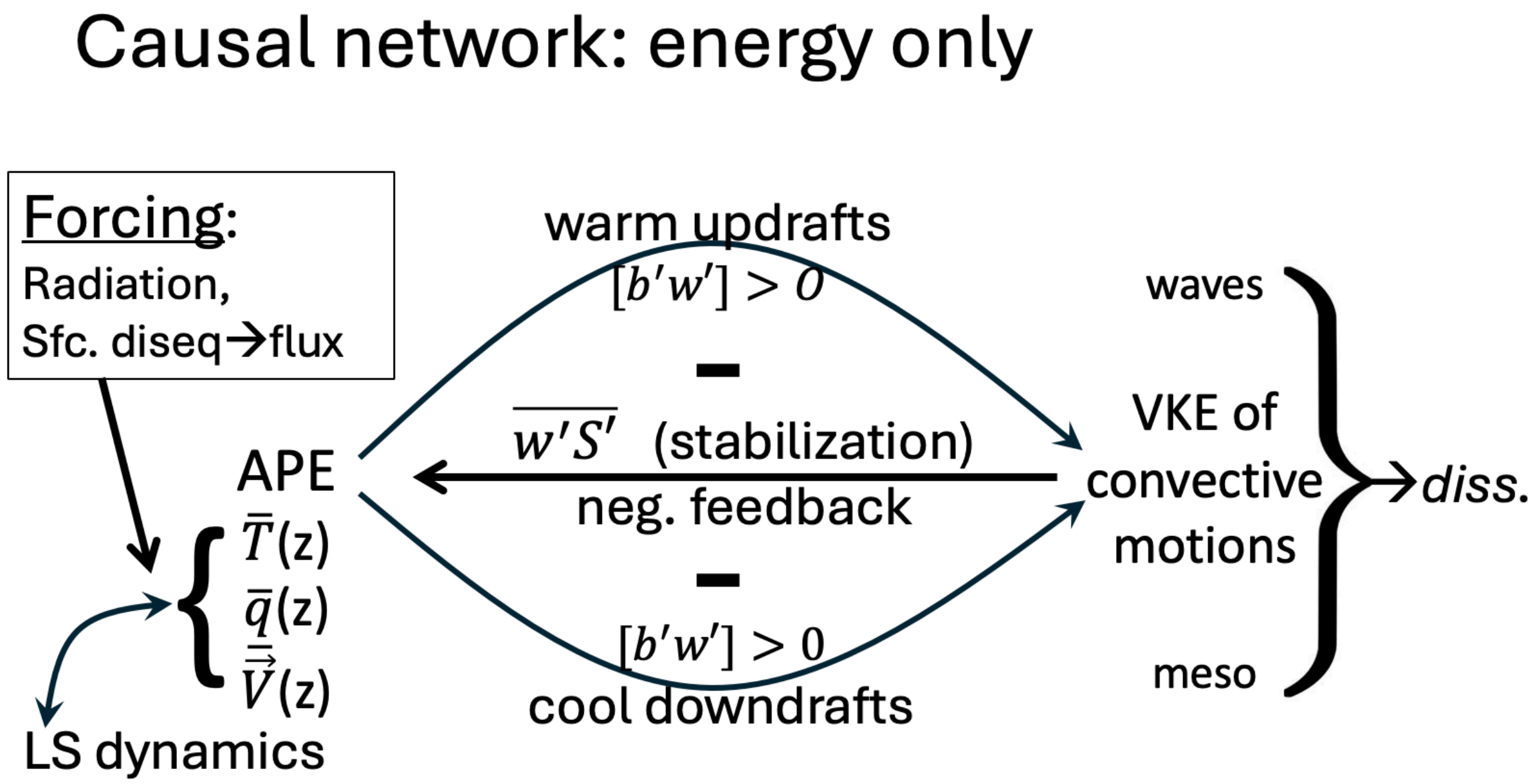

*Fig. 1: Causal network of a simple energetics view of vertical convection. Arrow heads show influence, and – indicates a central negative feedback loop that makes the system stable. Overbars are horizontal (e.g. LES domain) averages, primes are deviations from that. Brackets [b'w'] represent a mass-weighted volume integral, appearing in the KE budget of the Boussinesq or anelastic sets. Buoyancy b initially*

*drives vertical motion (VKE), but kinetic energy instantly spills into all 3 vector components, and to wave and mesoscale motions, before dissipation. Here S' is some conserved entropy, proportional to b' only in dry convection.*

Figure 1 invokes vertical kinetic energy VKE = ½ [w'w'], along with its familiar source term, the covariance of buoyancy and vertical velocity [w'b']. Although the code calls this quantity "buoyancy flux" (BF) following many authors, this is a misnomer since b is not a conserved variable in moist convection. Kinetic energy instantly spills into all 3 vector components, and to wave and mesoscale motions, before dissipation. The proximate driver for [w'b'] and thus motion is an "available" potential energy APE, whose definition from temperature T and specific humidity q is ambiguous for moist convection (e.g. Gertler et al. 2023). Stabilization happens through vertical flux of conserved thermodynamic quantities, shorthanded here as w'S'. Radiation and surface energy flux are shown as Forcing or destabilization: infrared cooling of air aloft (lessened by solar absorption in daytime), and surface fluxes driving near-surface air toward thermodynamic equilibrium with warm water. In this KE-based view of convection, the ratio [w'w']/[b'w'] is a key time scale, a residence time for eddy energy, about 10 minutes in the Cloud Botany simulations. For purposes where that time scale is considered negligible, a "hard convective adjustment" or mixed-layer scheme can simply set parcel b=0 and the q profile to well-mixed in a convecting layer at every instant. Larger-scale (LS) dynamics then feel the weight of that neutralized density profile. In this view convection is in equilibrium, consuming APE at the rate forcing generates it.

A "relaxed" or "soft" approach drives profiles toward a "quasi" equilibrium state with some time lag. For all the elaborations unleashed by the "quasi" prefix (reviewed in Yano and Plant 2011, 2020), the concept remains of adjustment of some univariate quantity toward an optimum (Arakawa 2004). Shear adds some complications, including an additional source of KE for convection which is ignored here. The non-uniqueness of moist APE or a "neutral" profile complicates analysis, but cracks open the door for patterns or configurations of convection to matter on larger scales. Through that crack pour interesting questions about cloud patterns and their ecology, as next-order elaborations to the fairly adequate first-order mixed-layer or adjustment ideas.

Just as entropy (related to information) is a conjugate variable to energy in thermodynamics, this paper is predicated on the idea that patterns (separate from energy; related to information) participate in a second set of important interactions in the moist convective process. Figure 2 extends Fig. 1 by indicating several ways that pattern information can interact with convection's KE budget. Reservoirs of KE other than the cellular convection itself (*meso, wave*) are part of what we mean by patterns.

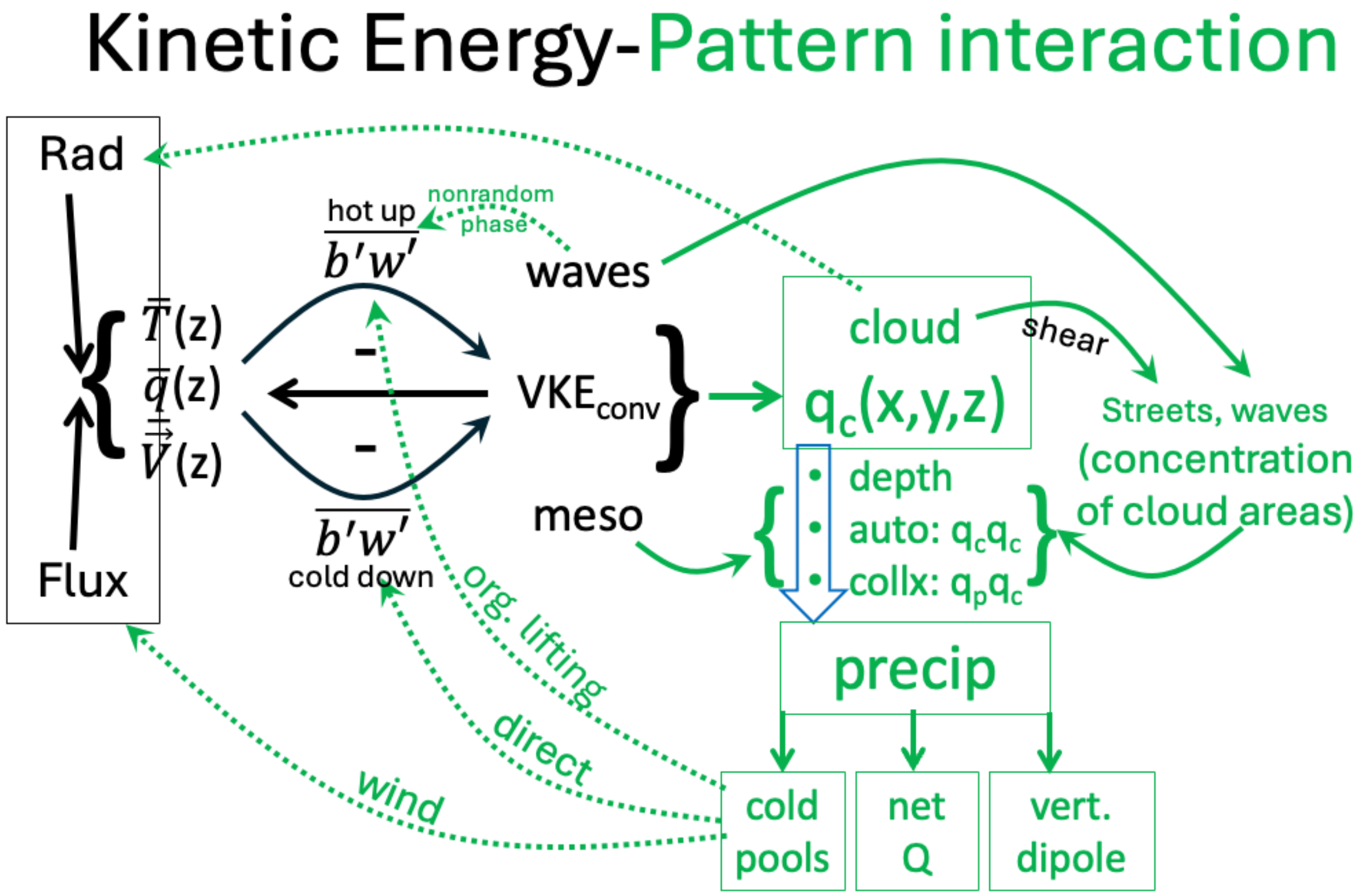

*Fig. 2: Larger directed causal network of vertical convection relationships, extending elements from Fig. 1 (black) with horizontal pattern related feedbacks (green). See text for more detailed process explanations.*

Metastability (conditional instability) means that non-convective KE can shape moist convection's [b'w'], either by initiating buoyant updrafts directly, or by modulating the inhibition energy required for cell-scale turbulence to do that job. Shear interactions, and especially precipitation once it forms (elaborated below), drive and sculpt mesoscale flow structures much larger than cell size and layer depth. Although internal waves are defined by quadrature between w and b, so [b'w'$_{wave}$] = 0, their crests vs. troughs present nascent convective drafts with different local environments, modulating their fates. Such phase coincidences (systematic deviations from randomness, evolving by natural selection) are indicated as "nonrandom phase" in Fig. 2.

Cloud effects are emphasized in Fig. 2, beginning from cloud water mass concentration $q_c$. That quantity is important both for its observability, and as a key nexus of causal paths toward precipitation, via the nonlinear (quadratic as indicated) autoconversion (auto) and collection (collx) production terms for precipitation water content $q_p$. Both processes are mediated by cloud depth, as well as by horizontal confinement or concentration of a given amount of cloud. Once formed, precipitation (after exploding via the collx term) drives evaporative cold pools. Sinking cold air is a direct KE source, and the resulting outflows are powerful mesoscale horizontal pattern makers at the surface. Cold pools trigger buoyant updrafts as a next KE-making setp. Net surface precipitation also implies a vertical dipole of latent heating Q with the positive part predominating (net Q). All of that affects internal waves and surface winds and fluxes, as well as driving "org. lifting" on scales larger than the individual parcel or simple cell scale, which may modulate and thus harvest (at their wavelength in Fourier space) buoyant PE→KE conversion via [b'w']. The arrows become too numerous to depict. Precipitation is so powerful that its history can be inferred from the patterns it induces in shallow cloud fields (Zuidema et al. 2012, Vogel et al. 2021, Haerter 2019, and many more).

The causal networks of Figs. 1-2 are conceptual postulates: webs of hypotheses about familiar qualitative pieces of knowledge, some learned from the analysis below, some conventional or purely logical. As a guide for quantitative analysis, they spotlight what to measure and suggest some (perhaps lagged) relationships to seek. Unfortunately, the existence of loops makes them less rigorously analyzable than the Directed Acyclic Graphs (DAGs) of formal causal study (e.g. McKenzie and Pearl 2019). For the present, we took Fig. 2 as a motivator to create a value-added depiction and dataset derived from the Cloud Botany simulation suite, offered in this article's repository. Its possibilities are far from exhausted by a few diagrams below, seeking to relate quantitative bulk energetics to patterns. This and the full dataset contain much more for some ambitious researcher with a more specific idea to pursue.

## b. Energy & pattern measures in Cloud Botany data

Internal quantities from the model outputs of the Cloud Botany set, computed for this article, include:

- Domain mass averaged [$w\theta_v$], converted to [b'w'] BF units, and its cumulative integral IBF
- Vertically averaged [ww], twice the VKE indicated in the causal graph
- HKE = ½ ([uu]+[vv]), an indicator of larger scales in the kinematic and wind fields
- Domain precipitation rate PR and its cumulative sum IP every 200 minutes
- Anisotropy and mesoscale heterogeneity measures of LWP(x,y,t) described below.

To reduce data volume, horizontal fields of vertically-integrated liquid water path (LWP) and precipitation rate were sub-sampled by a factor of 36 (6x6) horizontally, to 256x256 pixels from the original 1536x1536, and also 10x in time to every 200 minutes. Precipitation rate was then averaged over those subsampled pixels, while LWP was analyzed for its horizontal pattern information.

To distinguish purely pattern information from cloud-amount variations in LWP arrays, a Boolean image was created at each time in each array in which a True or 1 value indicates membership in the spatial upper decile of LWP (that is, the top 10% of pixels at that time). In arrays with less than 10% LWP>0 coverage, all pattern measures were set to zero. Each 256x256 Boolean array was then spatially coarse-box averaged over all powers of 2, yielding 7 floating-point spatial arrays sized from 128x128 to 2x2, each of which still sums to 0.1 by construction. After multiplication by 10, these become horizontal probability distributions (PDs, which must sum to 1). Specifically, each box value is the probability that a randomly selected pixel from that box would be in the top-$10^{th}$ percentile of LWP for the whole domain at that instant. These PDs are the basis for mesoscale information and anisotropy measures.

Information H is a scalar measure of PD nonuniformity, at each time, at each box scale. H is defined from entropy $S = -\Sigma\, p_i \log(p_i)$ for any PD = $\{p_i\}$ on any domain, with $p_i \log(p_i)=0$ taken when $p_i$ =0. S has units of *bits* if log is base 2, so a unitless or normalized information $H_n = (S_{max} – S)/S_{max}$ is constructed using the maximum entropy $S_{max}$ of a uniform PD on the same discrete domain. $S_{max}$ obtains when a PD = $\{p_i\}$ = 1/(number of boxes) everywhere. H is zero for a uniform PD (no pattern), 1 for a spike PD, and something in between for all other variable probability scenes. Entropy S is not very different from a probability variance $\Sigma\, p_i^2$, but information theory gives it elegant mathematical properties so it is preferred here. Some authors dislike the dependence of H on data resolution (number of boxes) and seek some extrapolation toward an asymptotically infinitesimal pixel size (Li et al. 2018), but here we value the ability to spotlight variations among 32x32 boxes, each about 5km wide, as representing "mesoscale" patterns. The $H_n$ spectrum for all 7 rebinning box sizes tends to vary together (not shown) so the 5km box size $H_{n,5}$ is chosen here.

Anisotropy A is a measure of elongated, oriented structures in cloud scenes. When a PD of updrafts is concentrated into half the area or collapses toward a single 1D band, that heterogeneity can perhaps foster precipitation development, for instance by increasing quadratic interaction terms $q_c q_c$ and $q_p q_c$. As an objective measure of anisotropy, a 2D Fourier transform is computed from the 256x256 upper-LWP-decile membership arrays. For anisotropic (spatially oriented) patterns, its squared amplitude (probability variance), averaged over the annulus of total wavenumbers 2-20, as a function of angle in 15-degree wedges, has a peak. A simple strength index for this peak is defined as the maximum divided by the mean, denoted as A. Values are rarely at the theoretical minimum of 1, and are never found to exceed 3.5. A-1 (where zero now means statistical isotropy) is shown in figures for clarity.

Figure 3 illustrates energy and pattern measures from simulation number 34, chosen for the diverse phenomena it illustrates. All 103 simulations are depicted thusly in the supplementary archive.

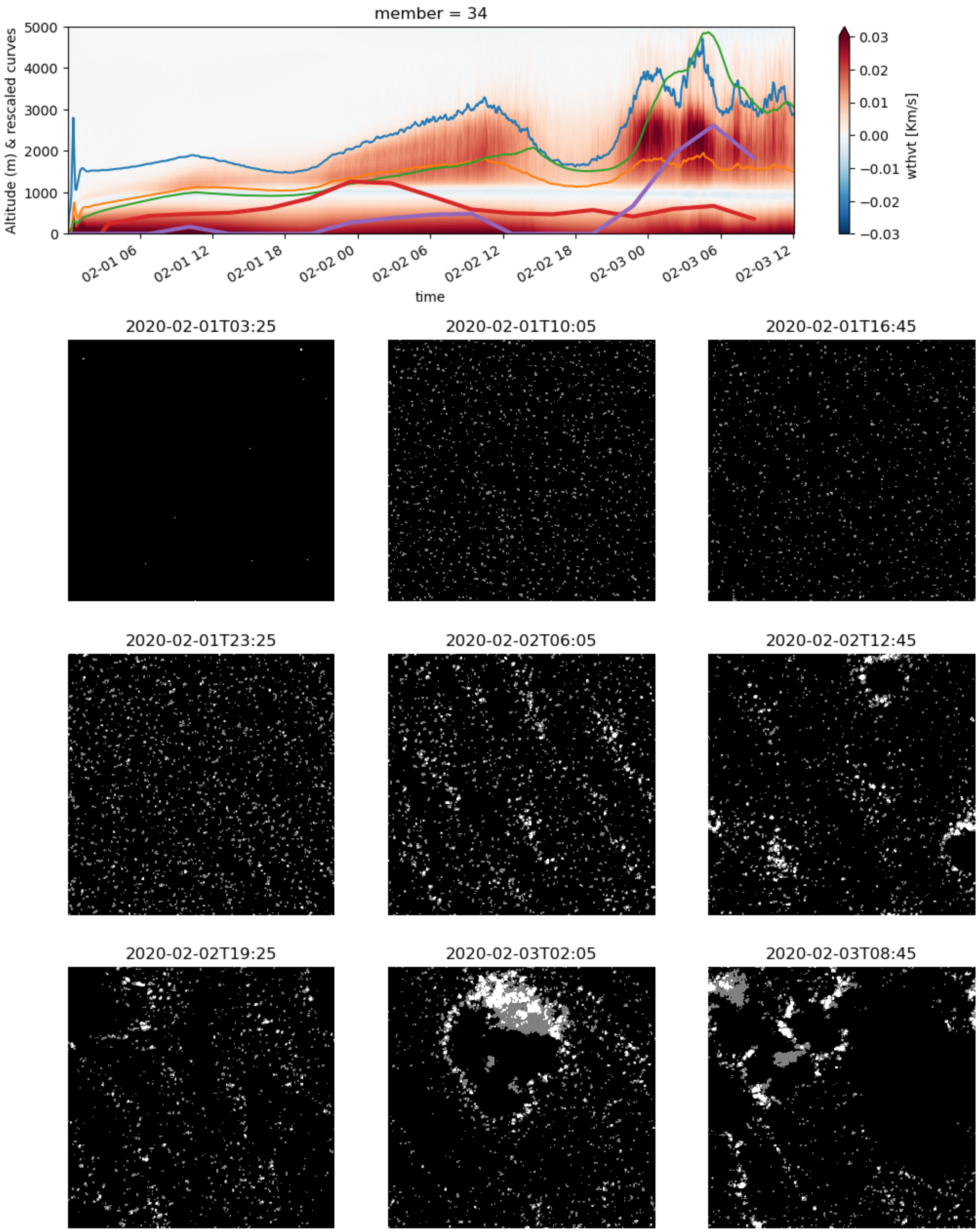

*Fig. 3: Time series of energy and pattern measures (top), and 9 image samples, from simulation 34 of the Cloud Botany set (Jansson et al. 2023). In the time series, the product* w $\theta_v$ *is red shading. Thin curves are kinetic energy budget terms: blue is mass-weighted vertical mean [b'w'] times 600s, orange is [VKE], and green is [HKE]/3, all rescaled so that 2000m altitude corresponds to 0.1 J/kg in VKE. Thick curves are pattern measures, only computed every 200 minutes: red is anisotropy 1000*(A-1), purple is 20000 times the normalized pattern information measure $H_n$ for 5km rebinning boxes. Cloud images are*

*156 km squares showing filled contours of LWP at levels 0.2 and 0.6 kg m$^{-2}$. A PDF document with the same figure format for all 103 simulations is in the supplementary materials.*

Three nights, with two intervening days of solar absorption, cause convection to have 3 active and 2 partly-suppressed epochs, seen most clearly in the cloud layer above 1000m altitude. Mass-averaged VKE (orange) is a bit over half as great as 10 minutes of [b'w'] production (blue), and varies nearly in phase as this short timescale would suggest. HKE (green) is about 3 times VKE for the first two nights, but rises to 6 times VKE and exhibits a 2-3 hour lag behind 2-3 hour pulses of convection on the third night, when convection is deeper and precipitation is active, seen by cloud rings and arcs in the images. That is measured also by >5km mesoscale pattern $H_{n,5}$ in the thick purple curve, similar to results by Lamaakel and Matheou (2022). Anisotropy in wavenumbers 2-20 (thick red) peaks early in the second night, near the third or fourth cloud images, but that banded concentration of cloud activity does not achieve precipitation in this case, perhaps since cloud depth is insufficient at that time.

From the dataset of 103 simulations, statistical relationships can be sought between pattern measures and energetics, as motivated by Fig. 2. Since the process of evolution is time-bound, scatter plots can miss the point, but are informative as context on a plot. Pattern observables ($H_n$, A) are chosen here as diagram coordinates, since the intellectual project is to learn to infer energetic processes and efficiencies as a function of satellite-detectable cloud patterns.

Figure 4a shows that vigorous VKE (green-yellow colors) is seen almost uniformly across all scenes, except for low values in the very uniform scenes (low $H_n$ at bottom), which correspond to early times in the simulations. If evolution's pace is set by eddy KE turnover times, rather than clock time, the cumulative [w'b'] called IBF in panel 6b is arguably a proper airmass "convective age" metric, and shows that the scenes near the bottom are indeed early in the convection's evolution, when it has processed little energy flux. The case from Fig. 3 with its second-night anisotropy peak and third-night $H_n$ spike with precipitation, is shown in a red path connecting larger dots, which (as they must be) are monotonically increasing in value in Fig. 4b. Finally, area-mean precipitation rate (Fig. 4c) is seen to be closely related to $H_n$, with little connection to the anisotropy measure.

The main lesson, perhaps unsurprising in retrospect, is that precipitation predominantly governs (and also benefits from) the development of pattern information. Its development in turn is strongly predicted by cloud depth (perusing the equivalent of Fig. 3 for all cases), which is controlled largely by thermodynamic factors (lapse rates of T and q in the initial conditions).

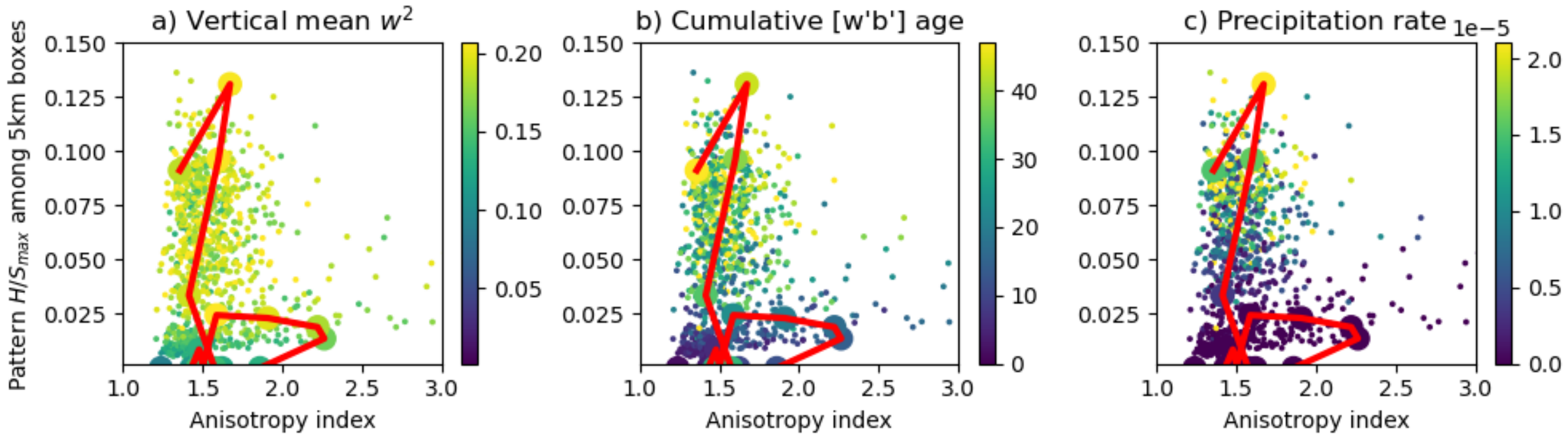

*Fig. 4: Case 34 from Fig. 3 (red path connecting large dots) overlaid on scatterplots of all 103 simulations at all available 200-minute subsampled times, on axes of the satellite-observable pattern measures anisotropy A and >5 km mesoscale cloud heterogeneity $H_{n,5}$. Colors indicate a) vertically integrated VKE, b) an airmass age measure (cumulative kinetic energy generation IBF), and c) precipitation rate at the time in question. All units are MKS.*

It is clear from perusal of the 103 version of Fig. 3 that cloud patterns do evolve with time, from uniform smooth initial conditions, in these simulations. Mesoscales develop gradually but systematically. The *how* of this length-scale growth (import of moisture by condensation heating) is detailed lucidly in Janssens et al. (2022). But does this suffice to explain *why*, if that question can be asked meaningfully? The conjecture that some kind of energy efficiency is increased by repeated selection, winnowing flow configurations from the simplest ("popcorn" convection) to more improbable and contingent multicellular structures, may be supported by observations (McCoy et al. 2025). Should we interpret evolution's imprint as predicting an increasing "efficiency" of VKE in producing its own source, [w'b']/[w'w'], or the longevity of VKE's lifetime, given by its inverse? Unfortunately, the secular trend of layer depth and wind-dependent surface fluxes and a slow daily radiation cycle affected by the clouds themselves make it impossible to disentangle the possible horizontal-pattern effects I initially sought. Model experiments by intervention, rather than merely output analyses, may be needed as in the "convective memory" literature (Colin et al. 2019, Daleu et al. 2020, Hwong et al. 2023) for instance.

One importance of convective organization to larger scales can be easily and clearly seen: overshoot of neutrality. Figure 5 shows that within the subcloud layer, positive stability of the mean can occur in precipitating scenes. Cold outflow pools at the surface spread to cover a large area, even as convection persists – in a subset of the domain which reliably finds itself by natural selection. This finding is unsurprising, but it is worth recalling that mean density is what the next-larger scales actually respond to, and that "neutrality" is not a hard limit of adjustment, even in the unsaturated boundary layer, when moist convection is present.

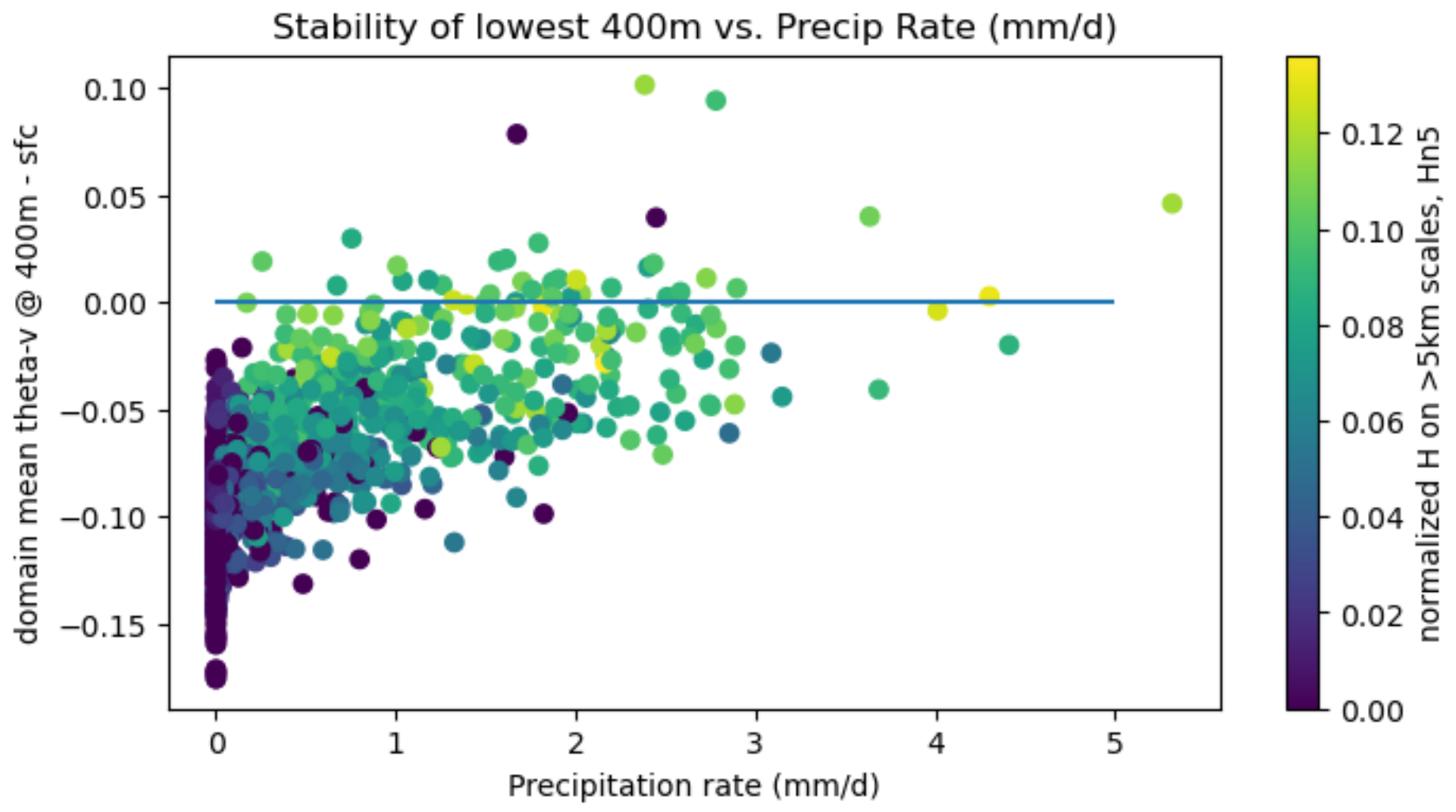

*Fig. 5: Scatterplot of lowest-400m stability (in virtual potential temperature K units) for all simulations, at the 200-minute intervals when LWP and precipitation were downloaded. Weaker*

*instability and even positive stability (overstabilization of the domain-mean profile by convection) is seen when precipitation and mesoscale (>5km) pattern information are large.*

Since precipitation has such a leading importance in convection pattern evolution, even for shallow moist convection, obscuring horizontal pattern evolution as a process per se, let us embrace that and jump to a simpler, more two-dimensional setting where the strong determinative trend of cloud depth has largely leveled out: deep convection.

# 3. Deep convection pattern evolution by cell-cell interaction

For deep convection too, in the absence of large-scale "forcing" imprints, the pattern of probability of new cell development depends substantially on precipitation from prior cells, both near and far in space and time. To explore the implications of this contingent evolution process, this section describes a two-step analysis to calibrate the probabilities in a stochastic model. Previous efforts of this kind include Hagos et al. (2018) or Cardoso-Bihlo et al. (2019), or the more deterministic considerations of Freitas et al. (2024).

Specifically, this section describes:

1. Empirical estimation of the near-field part of lagged cell-conditional probability (called a "kernel" for its boundedness), using infrared imagery over lowland South America; and then
2. Monte Carlo simulations of the iterated behavior of the conditional probability kernel from step 1.

## a. Satellite estimates of conditional cell frequency

To study evolution in patterns of deep convection, we define a discrete "cell" state (0 or 1) at every pixel location, and tally frequency of occurrence. Comparisons across years and between land and marine environments help spotlight unforced self-organization.

Specifically, three-hourly IR imagery during daytime were downloaded from the NOAA GridSat dataset (https://www.ncdc.noaa.gov/gridsat), at the same five time levels each day, with 0.07$^o$ resolution (approximated as '8km' at these low latitudes). Downloaded domains covered 20x20$^o$ lat-lon boxes just south of the equator (20S-Equator), over South America (SA; 40W-60W), and over the South Pacific (SP; 145W-165W. The sample used here includes 235 days in SA and 145 days in SP, spread over 20 years. For historical reasons, the sample is defined by days on which a Mesoscale Convective System (MCS, as defined by Chen et al. 2023) was identified to initiate somewhere in larger spatial boxes, at a specific local evening hour (23UTC in SA, 9UTC in SP). But that criterion is not important here except as a general wet-season indicator, as revealed by comparisons below with data on an equal number of days exactly 1 year later. That +1 year reference set has identical diurnal and seasonal sampling, but different weather, yet its general statistics are indistinguishable (not shown).

Defining convective cell entities on this 8km pixel mesh lets us tally information about space-time patterns of their frequency of occurrence. To define cells, IR temperature was very lightly smoothed (with the *scipy* library's *gaussian_filter* function using sigma=1 pixel), and then local minima were identified with that library's *minimum_filter* function. Minima whose T is colder than -40C are declared to be deep convective cells. These choices satisfy the eye (Fig. 6) and were made on that basis.

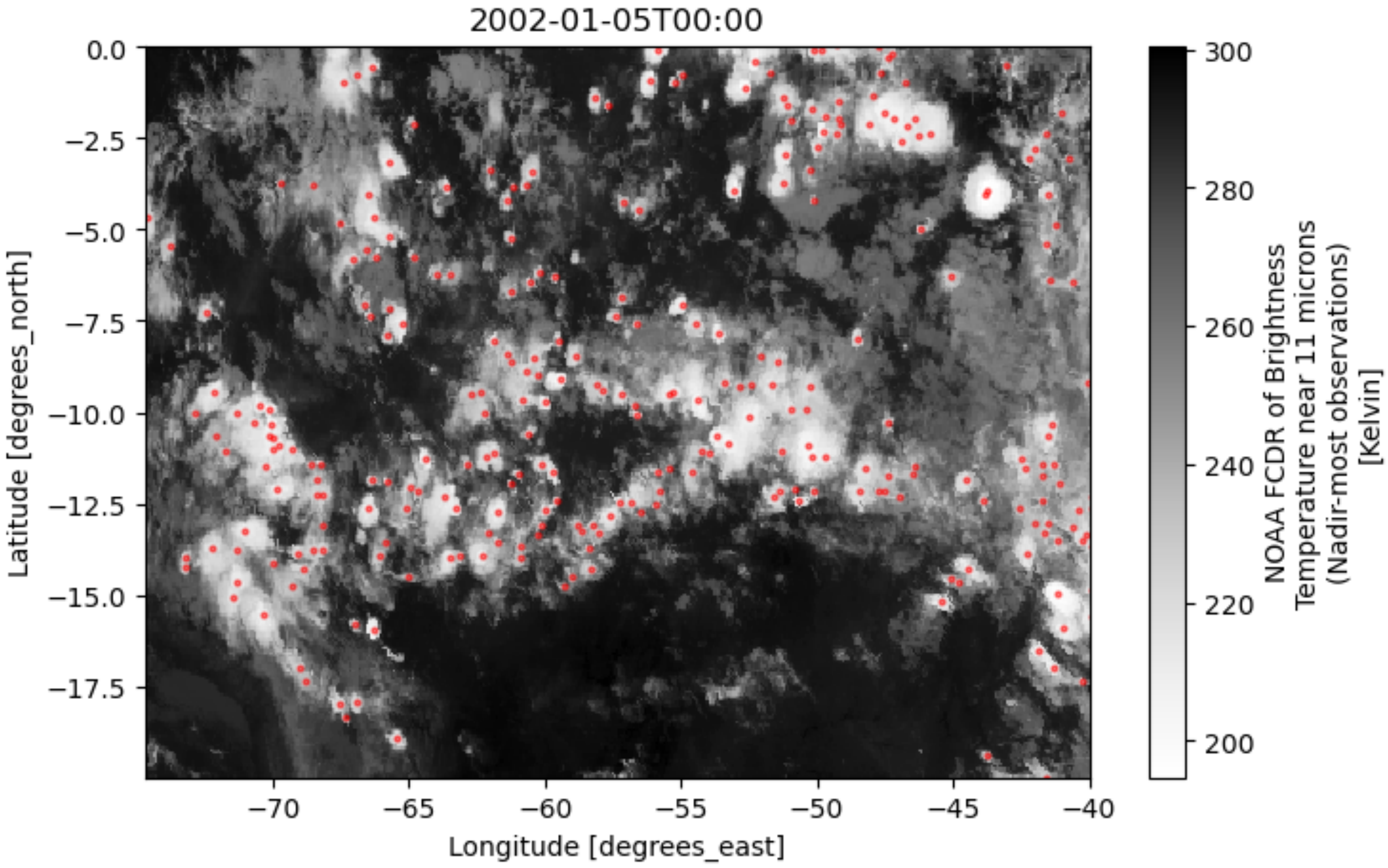

*Fig. 6: Cell identifications (red dots) as local minima with T < -40C in lightly smoothed IR imagery, on an example image over South America near 00Z on 2002-01-05 (8pm Local Time at 60W).*

To estimate how the *instantaneous* frequency of cell occurrence varies with distance from another cell, the 235 SA days (x2 including the +1 year set; and separately the 145x2 SP days) were pooled for analysis as a function of local hour. First, each Boolean cell array was zero-padded around the edges, so that no false adjacency signals are introduced in a loop which cyclically rolls the array to be centered on every available cell. The sum of all those cell-centered arrays was divided by its spatial horizontal sum, meeting the defining characteristic of a horizontal PD. The resulting frequency is notated like a horizontal PD, denoted $p_{cell}$ (x,y,h | cell@0,0,h), where x and y are distances from the central point (an average reference cell's location). This PD notation is slightly confusing, since the origin value is not 1; rather the horizontal sum of the whole pattern is 1, akin to percolation theory's *probability of occupation* at each site. Padding contributes a mild artificial decay with distance, by diluting the frequency map for cells near the edges of the 20x20$^{o}$ lat-lon box. While larger boxes would reduce this edge effect, that would incur more geographical heterogeneities. In the end, the 20x20 degree area of quasi-uniform lowland tropical South America was the compromise accepted here. Patterns are kept separately for each clock hour h. Over South America, times are most of an hour after [11, 14, 17, 20, 23] UTC; subtract 4 for approximate local time at 60W. Over the SPCZ, similar *local* hours are used, 15 time zones away.

Results for South America (SA) are shown along the diagonal in Fig. 7 in blue-yellow colors. The displayed quantity is $\log_2$ of $p_{cell}$ (x,y,h | cell@0,0,h), in 1000km x1000km cell-centered squares. Besides the main result -- a near-circular local enhancement on a scale of about 100 km -- artifacts include a faint crosshair from edge effects, a small ring of zero near the origin where the Gaussian smoother makes local minima impossible, and sampling speckle. At 14UTC (10am local time) the mesoscale central-region frequency enhancement is greater (brightest yellows are in upper-left panel). This reflects the fact that deep convective cells at 10am mainly occur when a previously developed mesoscale storm was active overnight, a somewhat rare occurrence; 10am is otherwise a low-cell-frequency time of day before deep convection commences in earnest. This pre-existing storm signature will be seen again below over the maritime SP area, where night does not interrupt convection's evolution.

To estimate a *time-offset* conditional frequency of cell occurrence, as a function of distance from any given cell, the same padding and rolling-to-center algorithm was used. Here an inner loop ran over only the cells in a "base" clock hour b. For each cell-centering operation in the loop, the entire stack of identically-rolled cell arrays at all clock hours h was summed and normalized. The result is a function of b and h denoted $p_{cell}$ (x,y,h | cell@0,0,b). Off diagonal in Fig. 7 (the blue-red color scale) are 1000x1000 km arrays of $\log_2$ of

$$\frac{p_{cell}(x, y, h \mid cell@(0,0, b))}{p_{cell}(x, y, h \mid cell@(0,0, b\ 1\ year\ away))}$$

where h is the UTC clock hour (column in the figure) and b is a base UTC clock hour (row in the figure). The denominator is constructed from the 235 cell arrays 1 year offset from cell locations at base hour b, for each day in the dataset.

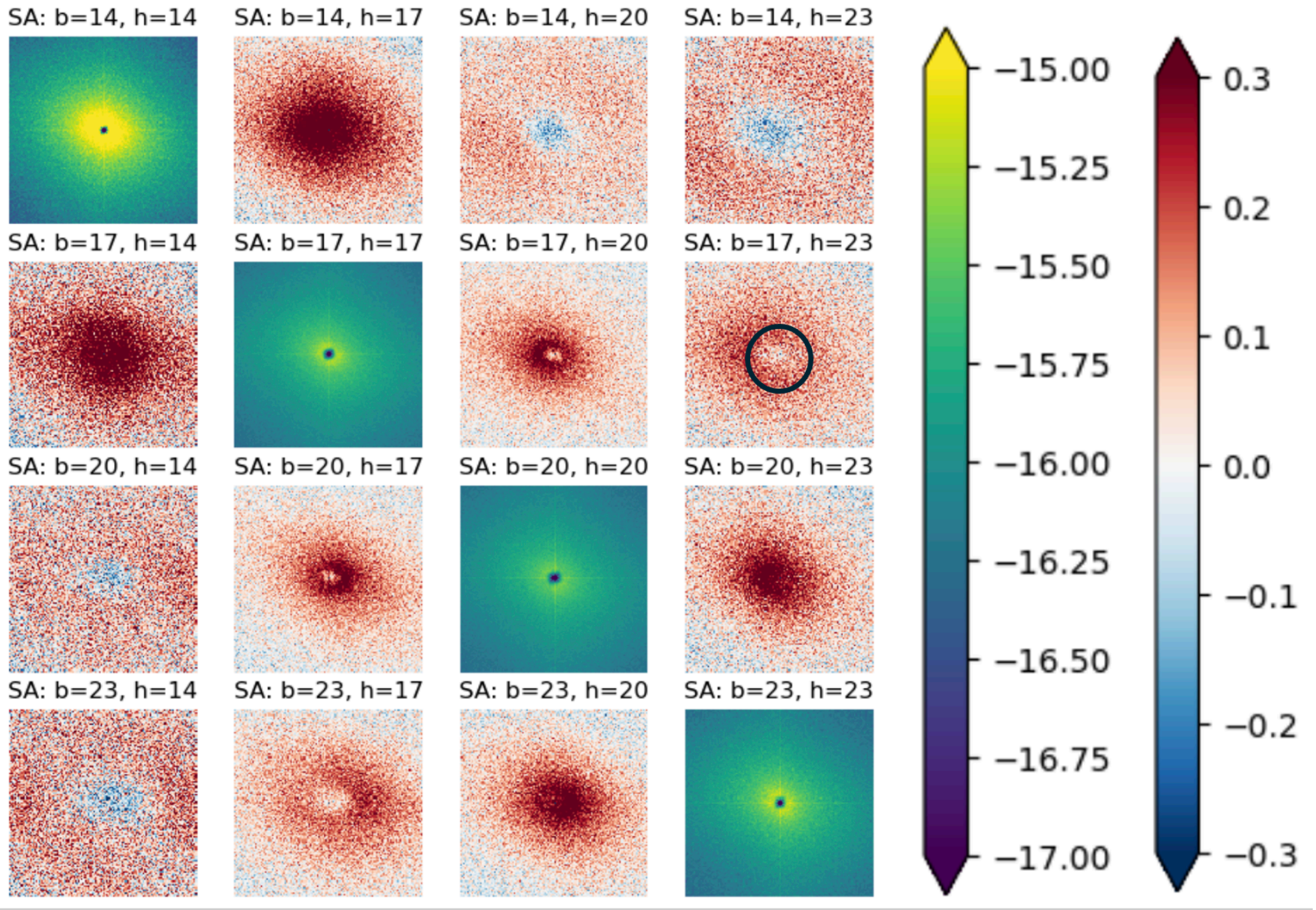

*Fig. 7. Relative frequency composites through late morning-evening local hours, for 1000 km square areas centered on each cell location in the SA (South America) dataset. On the diagonals (blue-green-yellow scale, positive definite) are instantaneous $log_2$ of $p_{cell}(x,y,h)$ for clock hours h. Off diagonal images (blue-red scale) depict differences of $log_2$ of conditional frequency of a cell at clock hour h, given a cell at the center at base hour b* ***one year later or earlier****. Base times b and display times h are UTC hours; for LST subtract 4 roughly. The black circle with diameter 333km after 6 hours indicates a ring expansion speed of 8 m/s.*

Causal interpretation of the conditional frequency pattern as cell-*driven* is aided by using the denominator from a year away (an independent weather sample). Any conditional patterns inherited from climatological structure, such as geographically anchored features by season and hour, are normalized out. Also eliminated is the identical structure of the cell-finder artifacts near the central point. No difference was seen between composites based on the cells in MCS-related reference days which drove the original 235-day sample selection, and in the reference days one year later, so the two offsets (forward & backward by a year) are pooled here to double the sample size of this result. Since the log of a ratio is a difference of logs, values can be negative or positive.

Again the morning results (left column and upper row) are distinctive, reflecting the influence of relatively uncommon days with cells at 14 UTC (10am), often in pre-existing overnight mesoscale storms lasting until about 1pm. Their local impact later in the afternoon is a *reduced* frequency of new cells (blue central feature in upper-right panel), perhaps because cloud cover reduces insolation. The lower-left panel is similar to the upper-right, with a blue central feature of about 200 km scale, but it addresses a slightly different question: What *morning* (b=14UTC) cell frequency pattern is associated with an evening (h=23UTC) cell at some location? The answer is related to the same phenomenon: evening cells are more likely after mornings which do *not* have earlier cells from mesoscale storms active at that location. These results resemble the maritime case (discussed below).

No diagonal symmetry is enforced across Fig. 7 mathematically, but the iterated process of cell-cell interaction does seem to produce it. One subtle asymmetry is more speckle in the lower-left than upper-right, for instance comparing b=14, h=20 vs. b=20, h=14. Forward cell-conditional probability may be expected to be more deterministic than backward: while some cells grow spontaneously and randomly, without prior cell effects, every cell's forward impact is definite and deterministic. This conjecture or inference could be tested with outputs from the next section's iterated kernel model.

An expanding-ring pattern is seen through the afternoon hours (1-4-7 pm local), columns 3-4 in the second row of panels. A cell at the origin at 17 UTC is followed by an expanding red ring of enhanced occurrence frequency at 20 and 23 UTC, reflecting new-cell triggering effects such as internal waves and gust fronts at the edges of expanding precipitation-induced cold pools. The speed of expansion seems to exceed the 8 m/s represented by the black ring, suggesting internal waves may be a prominent component. A blue core indicates cell reductions (local stabilization), owing perhaps to that cold-pool air and/or surface shading by clouds. A modest east-west anisotropy indicates a mild westward drift velocity, and/or the sun's motion in the sky. By 23UTC (e.g. b=20, h=23) the blue core has largely disappeared, perhaps indicating that any 6-hour-old mesoscale systems lasting past sunset are starting to become generalized cell-enhancers, akin to the mature systems of the morning (upper-left) and oceanic SA samples to be examined next.

Figure 8 shows the same diagram for the South Pacific (SP) maritime sample. The diagonal panels show strong mesoscale instantaneous enhancements at all hours, indicative of cell frequency being driven by sizable (100s of km) mesoscale convective storms, similar to mornings over SA. Since this area is climatologically less convectively active than the SA box, the cell-conditional enhancements are stronger (more saturated colors). Also, the sample is smaller (145 days) so the patterns are noisier. Red blotches off the diagonal indicate that mature storms at all hours have typical duration > 6 hours. In this region, meso-synoptic enhancements of deep convection often involve trailing fronts from westerlies to the south, lending a NW-SE orientation to the composite structure. These all-similar and positive mesoscale (~500 km) conditional frequency enhancements seem to indicate a climax ecology of mature multicellular storms governing cell development at all hours of the day.

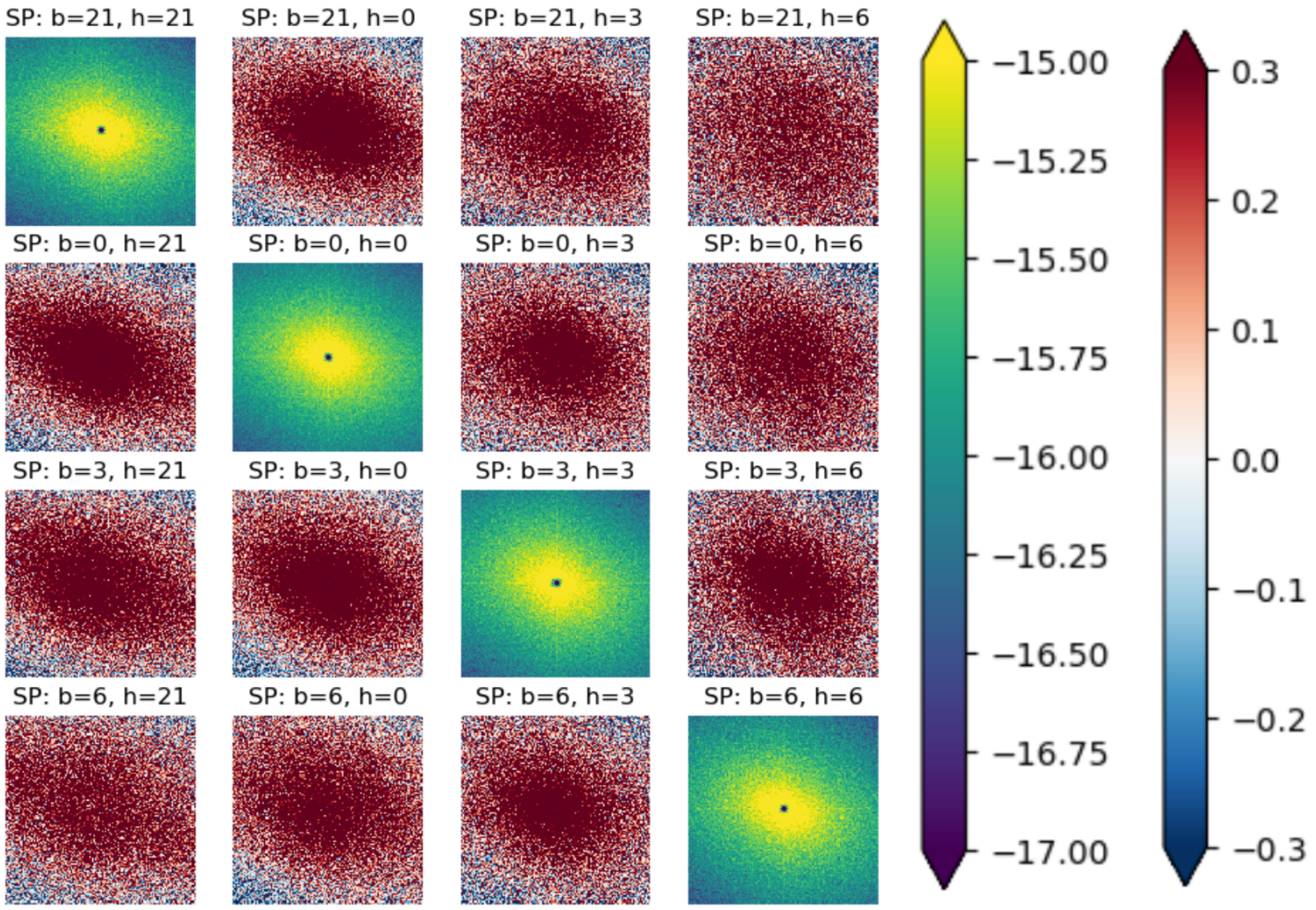

*Fig. 8. As in Fig. 7 but for the SP (South Pacific) sample. Base times b and display times h are UTC hours; for LST add 14 or subtract 10 roughly.*

## b. Iterated conditional probability kernels: a model of cell ecology

The conditional frequency pattern of Fig. 7, an expanding positive ring with negative central core, was idealized as a conditional probability kernel for Monte Carlo simulations. Iterated space-time conditional probability models, sometimes called stochastic cellular automata (CA, Crommelin 2016), or Conditional Markov Chains (CMC, Dorrestijn et al. 2016), or "game of life" exercises (Wolfram 2002; see rules 18 or 150 in the fascinating comprehensive atlas.wolfram.com) from which the "Game of Cloud” in Kruse et al. (2025) draws its name. Such iterated probability models are used operationally for simulating correlated or coherent structures for deep convection process parameterization (Bengtsson et al. 2011, Bengtsson et al. 2021, and intervening works). Figure 9 shows an east-west time-longitude section through a plausible kernel probability factor, based on the results implied by Fig. 7. This is reminiscent of Fig. 5 of Daleu et al. (2020) showing memory by scale spanning hours, broader than the nearest-neighbor statistics of Khouider (2014).

The kernel here is 21x21 pixels in spatial dimension, over 5 time levels (cell lifetimes), here labeled ‘hours’ based on a diurnal cycle of 24 time steps that can be multiplied by the model’s uniform base probability, p = 0.01 x ($1 + D \sin(2\pi t/24)$ ) with D the diurnal amplitude. An important aspect for the emergence of large-scale structure is any overall positive bias in the kernel’s product over the 5 times (that is, in the sum of the log-kernel over the 5 times). That sum is carefully set to zero here for clarity, although a space-time correlation of the positive parts of the kernel (corresponding to a slantwise sum through the log-p cube) can still emerge as we shall see. A slight left-right kernel asymmetry has also been introduced, multiplying values left of center by 1.05 and values right of center by 0.95.

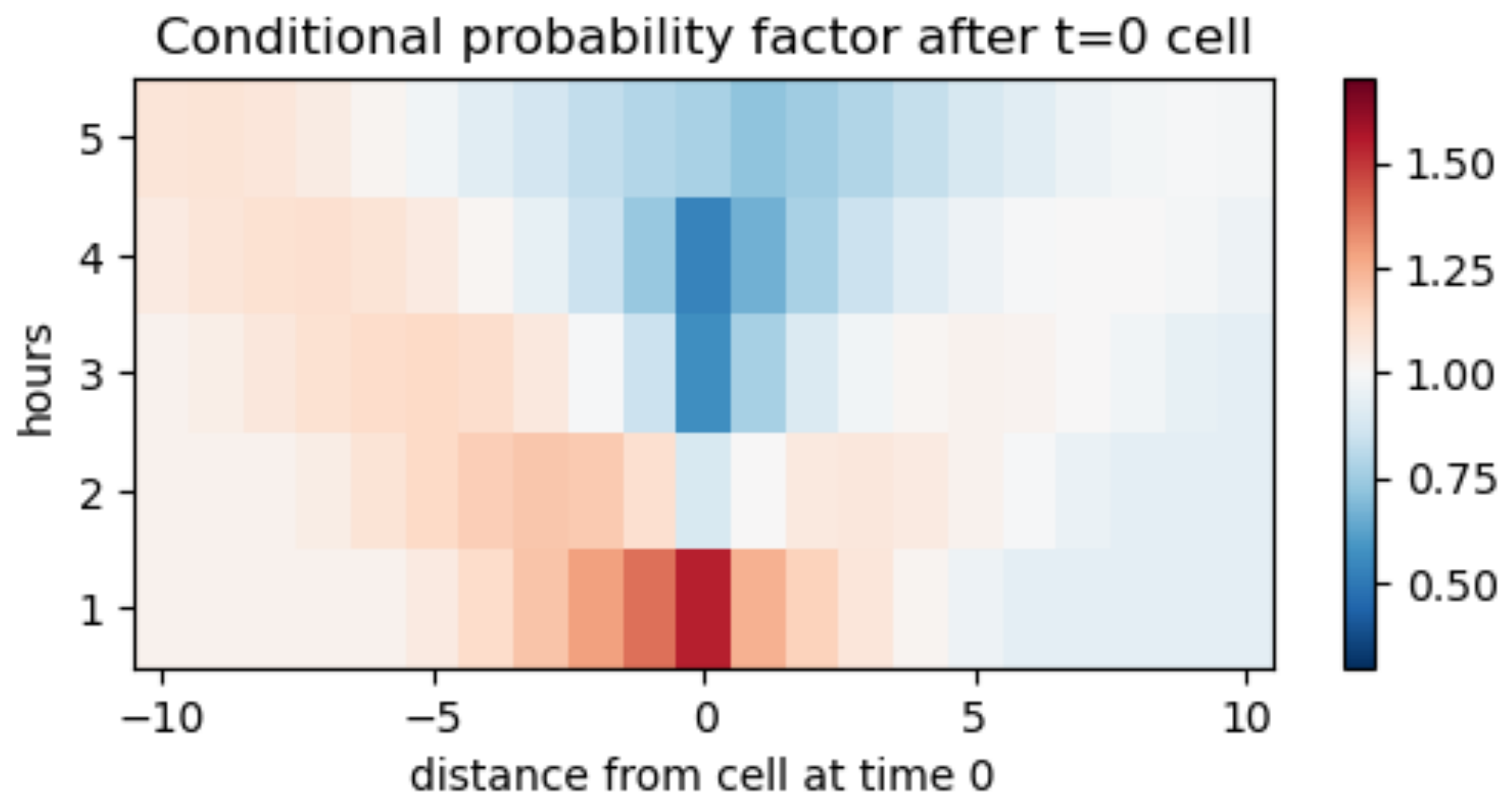

*Fig. 9. Expanding-ring conditional probability factor, called a ‘kernel’ for its bounded size. If 10 pixels over 5 “hours” here matched the 8 m/s speed of the black ring in Fig. 7, the implied pixel size is 14 km.*

Results of iterating this kernel for several tens of timesteps are shown in Fig. 10. At the initial time, a uniform background base cell probability of 0.01 on a 300x300 domain produces an expectation value of 900 cells according to a random number generator. Each cell then generates an expanding-ring conditional probability factor in its vicinity for the next 5 times, given by Fig. 9. At each subsequent time step, the uniform background base probability of 0.01 is multiplied by all these conditional factors from the prior 5 generations of cells. The resulting total probability map is then rescaled to have a mean of 0.01, to continue producing about 900 cells at each time step. With this rescaling, the cell population remains stable. Although the kernel is neutral (the sum of the log of its values is zero), factors exceeding 1 can multiply constructively in correlated structures, making the summed probability exceed 0.01. When we rescale in this way, cell probability in quiescent areas is reduced by those correlated structures. For this reason, there do exist configurations (squall lines as we shall see) that can *actively* suppress random cells elsewhere, as in simulations by Tsai and Mapes (2025) for instance.

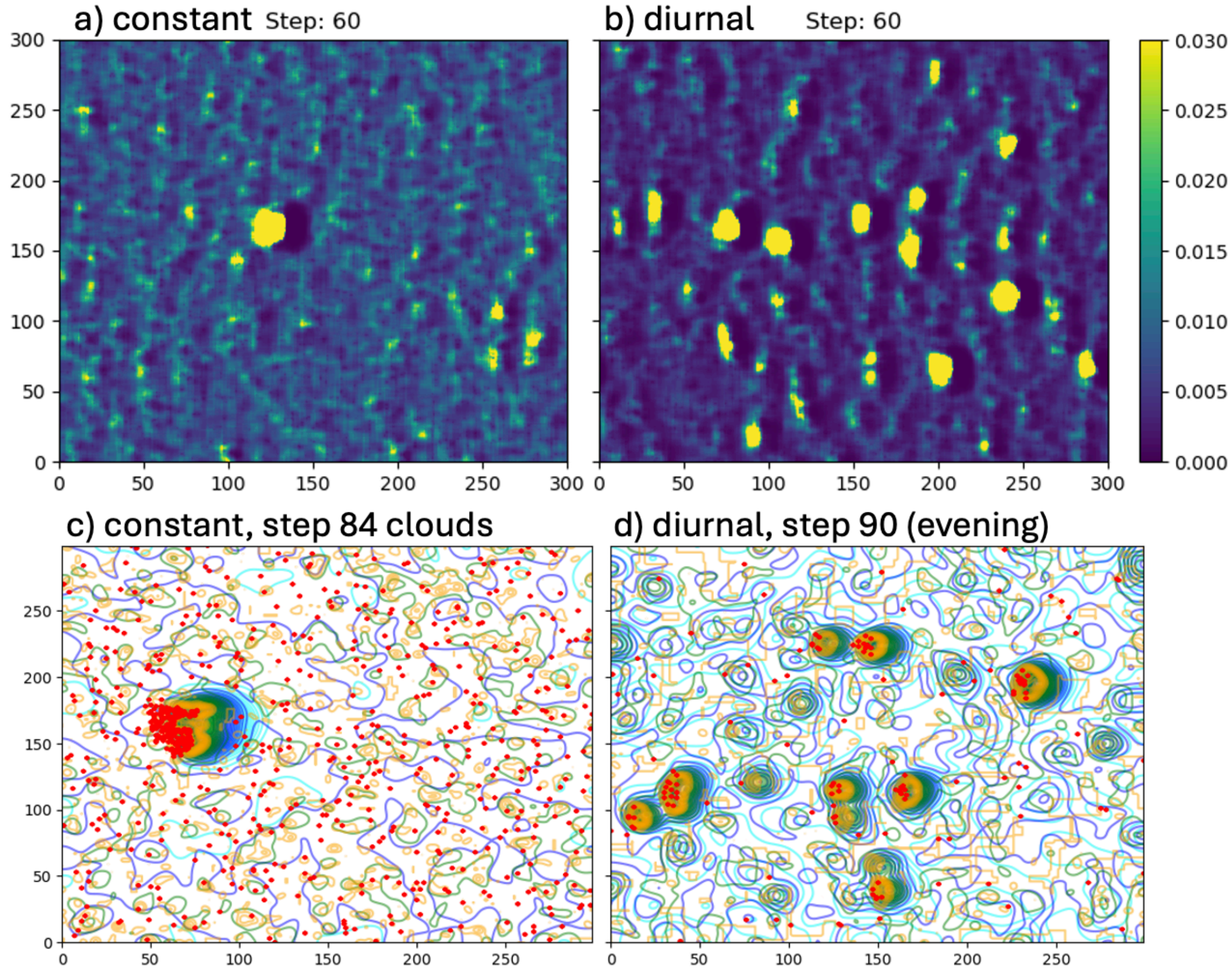

*10. 10. Top row: probability fields at hour 60, with diurnal amplitude of its specified horizontal mean of a) 0 and b) 1. Bottom row: c) hour 84 from amplitude=0 and d) hour 90 from amplitude=1 presented as pseudo convective-stratiform radar presentations (see text). All results are from an iterated probability Monte Carlo simulation using the kernel of Fig. 9; animations in supplementary notebook. A pixel size of 14km would make the kernel correspond to the speed of the black ring of expanding probability in Fig. 7.*

Probability is the top row of Fig. 10. The pseudo-radar depiction of the most-recent cell initiation time history array (lower panels) needs explaining. In a loop over each value in the set $\{age_i\}$ = {9,7,5,3,1} hours, the Boolean 0/1 array of the most recent cell initiation at a pixel being within $age_i/2$ of $age_i$ is constructed, and smoothed to a scale of $age_i$ pixels using scipy's gaussian_filter. The physical idea behind this is that older cells have produced anvil clouds which spread and smear with time, the "particle fountain" model of Yuter and Houze (1995). Contours of five colors {'cyan','blue','green','orange','red'} mapped to $\{age_i\}$ were created in succession, half-transparent, yielding something meant to resemble a radar depiction of spreading stratiform cloud and precipitation from several hours of recent cell activity before the present cells (which are red dots).

How long did it take this iterated-kernel system to discover its affordance of a special "organized" configuration of a squall, through evolutionary trial-and-error? In the case with constant background base p = 0.01 on a 300x300 domain (left column of Fig. 10), the first squall (a west-east probability couplet strong enough to suppress p elsewhere) appears only after about 50 "hours" or time steps (900x50 cell lifetimes). Once formed, the squall strengthens through hour 60 (shown at upper left), and lengthens in the north-south direction through t=84 (lower left panel). In the diurnal case with D=1*,* a doubled probability during "daytime" facilitates daily emergence within 5-10 time steps of about ten of these persistent multicellular structures (right panels), indicating a steep nonlinear dependence. Perhaps this dependence is quadratic (as in the *auto* and *coll* terms in Fig. 2)? At any rate, they can persist well into the "night" as base p plunges, like real MCSs. With enhanced probabilities of formation but clipped lifetimes, these diurnal squalls tend to stay near kernel scale (21 pixels), rather than elongating which takes a long time (four days to reach panel c). Animations of these cases may be viewed in the nbviewer supplementary link, and readers can easily run the pure Python model in a few seconds on a modern laptop, tweaking parameters and exploring consequences.

Output data from iterated-kernel exercises could be subjected to other analyses, such as attempting to recover a known kernel from its highly iterated statistical outputs -- a nontrivial matter, especially deep into a highly evolved situation, long beyond uniform initial conditions (Craiu and Lee 2006, Tulich and Mapes 2008). These activities are beyond the present scope, but ripe for effort, perhaps bringing useful empiricism to operational CA models (Bengtsson et al. 2011, 2021), which would then allow exploration of the larger-scale implications of mesoscale evolutionary kernels. Those impacts range from long-timescale indeterminism and persistence, to large spatial correlation scales, to overshoots of "neutral" mean thermodynamic conditions as in Fig. 5. Might such questions rejuvenate a sense of purpose in the descriptive enterprise of tallying "MCS" shapes and sizes (Feng et al. 2025), or bring new interpretations to analytic statistical models of clustering and percolation (Peters et al. 2009, Craig and Mack 2014, Ahmed and Neelin 2019, Hunt et al. 2020, Li et al. 2021).

# 4. Strategy sketch for evolutionary theory

Simulations like the above, perhaps with more elaborate or physical "cells" and their kernel of implications for the local environment (Kuo and Neelin 2025, Igel et al. 2025), could be a way to enumerate the size of a meaningful configuration space as a framework for evolutionary inference. The inverse of volume in such a configuration space defines the (very low) naïve probabilities of organized configurations emerging spontaneously. A Darwinian "fitness" function defined over that abstract space has gradients that express how strongly (per time step or generation) selection bias can improve on that naïve probability. That expression is a forward mathematical model of evolution, but we want the inverse: Can we infer or learn about the size of a meaningful or natural configuration space and its fitness and efficiency gradients from data on the enhanced frequency of occurrence of naïvely unlikely configurations? That project could be undertaken with data such as the vast abundance of satellite imagery.

If interactions are weak, configurations compete to exploit a common environmental supply of energy (instability). In that case, fitness may be a single time-independent function of configuration space, a fitness “landscape”. The weak-interactions case is assumed in *plume-ensemble* cumulus parameterizations for instance, where a 1-dimensional configuration space (plume width, corresponding to lateral entrainment rate, and thus to buoyancy top altitude) has a tilted fitness landscape inexorably favoring the larger. To make such schemes viable, a handcrafted “critical cloud work function” penalty spectrum must be tuned to handicap the competition, allowing inefficient shallow cumuli to exist in adequate proportion (Lord et al. 1983). Another longtime challenge in such schemes is to somehow delay the daily development of the more-efficient wider plumes, for instance with an organization parameterization (Mapes and Neale 2011, Park 2014, Zhang et al. 2025).

With stronger interactions in complex ecologies, fitness is a more daunting “seascape” of mutual contingencies (Cairns et al. 2022). Allowing for multicellular configurations also takes the evolutionary concept to a next level of generality. If it were true that clouds consumed each other while retaining identity, rather than consuming environmental instability with varying efficiency, “predator-prey” models from ecology could be fitted (as reviewed in Chen et al. 2025). However this approach may be more of a tool-borrowing exercise, rooted in a loose analogy and convenient observables (clouds), rather than a deep consideration of how evolutionary selection shapes convective configurations over time.

Coalitional strategies in game theory (Newton 2019) are another promising conceptual starting point. Multi-cell coalitions can be viewed as an additional ecological category, analogous more to a functional *guild* or *trophic level* than a too-specific *species* in biological life. Coalitions, benefitting from new efficiencies but perhaps vulnerable to disruption, compete for environmental energy against each other -- but also against simple, relentlessly spawned single-cell configurations. Applying these ideas hinges on first defining the term “configuration” and the space of its possibilities.

Configuration space could be built around a basis set of orthogonal wavelets (Yano and Plant 2025), contrived to add up to the total wind variance (kinetic energy), for instance in LES output data. Fourier analysis could also suffice. With such a complete but arbitrary basis, coalitions (weighted sums) are everything: even an isolated updraft (like a delta function) is a sum over all Fourier wavenumbers for instance. Phase relations contain the essential information, as randomizing phase from a delta function yields white noise, conceptually the polar opposite of a coherent structure.

Evolutionary selection would then play out in this cryptic space of noncompact sets of weights. While this seems daunting, the same thing happens in biology, where *pleiotropy* is at the root of life: genes-to-traits is a many-to-many mapping. Traits-to-fitness is another many-input mapping, making the whole thing entirely obscure to simple perusal. And yet evolution can and does occur meaningfully, including the evolution of cooperative coalitions (Sudakow et al. 2024). The only requirement for evolution is that causal connections are *robust*, not that they are compact for our brains to trace; pleiotropy is a puzzle, with robust aspects that can seem “surprising” (Reinitz et al. 2023). A mature field of bioinformatics is learning to see across the high-dimensional abysses of combinatoric complexity, and machine-learning discovery engines are becoming available for many applications.

# 5. Conclusions

Cloudy convection, Earth's most dynamic subsystem short of life, is extremely well observed and has important impacts. This makes it a compelling phenomenon for evolutionary study. An intimate, relentless, deterministic body force – gravity, buoyancy – makes it almost tautological to say that a convecting fluid will increasingly favor more efficient or "fit" configurations with time. Toy examples like Fig. 10 suffice to show that those better configurations can be so improbable (requiring 40,000 cell lifetimes in Fig. 10 for instance) that they may not be instantly or even inevitably discovered by the flow every day. The landscape between random (facile) and optimal (equilibrium) is large and fruitful, threaded with historically contingent and non-unique pathways up or down gradients that drive selection locally. This vast landscape, evocatively illustrated in Fig. 8 of Farmer (2024), is a large field for "play" (Burguillo 2013) with worldly importance.

The vertical dimension and invisible branches of convecting flow make cloud observations rather roundabout for quantitative study. Precise simulations help fill the gap, but still need conceptualized data processing to be interpretated. Here I postulated that the fitness or efficiency of flow configurations can be usefully measured in KE budget terms (Fig. 1), although the conjugate space of patterns or configurations (Fig. 2) has no such natural measure. Information H and an anisotropy coefficient A were examined here, but are somewhat arbitrary and scale dependent. Cloud depth (governed separately from horizontal patterns), leading to precipitation, was the main causal pathway in the Cloud Botany analysis of section 3.

Deep convection, where precipitation is a given, offered a cleaner 2D case for evolutionary theory, with "cells" and their conditional probability kernels as the simplest coin of the realm. Iteration of Markovian nearest-neighbor kernels has a long history of study, and even that simplest case eludes analytic analysis, so this is a computer-age problem requiring a New Kind of Science (Wolfram 2002). With nonlocal space-time kernels in 2D computation is all we have (as in Figs. 9-10). Within a cell-based or cloud-blobs version of 'configurations', sensitivity to kernel and background is a rich field as sketched the prior section 4. Intervention experiments, especially sudden changes that induce an ecological succession, can be informative at that level of description, in solvers of rigorous conservation laws in 3D (Tsai and Mapes 2025). But to go much further, "configuration space" will have to be rethought more deeply, and perhaps inferred in an inverse way from tallying the statistics of outcomes.

What is at stake in developing an evolutionary theory of organization? Larger-scale (planetary) convection has valuable predictability which depends on the bulk density of air columns. That in turn depends on convective patterns (Stein et al. 2017) which evolve over hours or even days, more general than the facile subcloud case of Fig. 5, on which gravity then acts to shape larger scale flow KE.

. Condensing (vs. not, or now vs. later) an aliquot of vapor causes a severalfold change in terms of the bulk density of an airmass. Importances include:

1. Understanding the evolutionary process could help us design better parameterizations. These could include refinements on crude deterministic mechanisms (Mapes and Neale 2011, Park 2014, Zhang et al. 2025), and better calibrations of stochastic CA-based schemes like

Bengtsson et al (2022), as in the project of Hagos et al. 2018, Cardoso-Bilho et al. 2019 and others. Rather long correlation lengths are found to be needed in such schemes, so perhaps an evolutionary perspective helps explain the origins of these long scales.

2. Initialization is the key to predictability, yet one of our largest and most informative datasets about the atmosphere (cloud imagery) is sidelined from the data assimilation process except for wind estimation. If pattern information has value, might it be discovered agnostically with convolutional networks, either as a pattern-to-profile retrieval (imagers-to-sounders), or in a blended functional discovery process like Shamekh et al. (2024)?

Finally, impacts models which bypass atmospheric dynamics such as hydrologic clustering (Ahmed and Neelin 2019) or percolation (Hunt et al. 2020) might be usefully informed by an appreciation of the evolutionary processes upstream of them. For instance, if slowing global circulation makes airmass lifetimes longer, might the statistics of extremes or cyclones be reasoned in terms of longer times for evolutionary selection to play out, with lifetime and evolution rate properly rescaled?

Denial of evolutionary possibilities by forbidding or excluding mesoscales offers another oblique glimpse of what is at stake. The enforced spectral gap in super-parameterization models (e.g. Pritchard et al. 2014, Tulich 2015, Jenney et al. 2025) offers a clean laboratory to measure and interpret how different-sized periodic convection-resolving domains may rectify into performance issues, including determinism, time lags, and extremal statistics. This framing lends more scientific meaning to domain-size experiments that have largely been viewed as mere computational engineering tests. Scale-aware stochastic parameterizations could be devised to fill the variable-sized spectral gap, perhaps requiring surprisingly long correlation and delay scales in space and time (“Memory”, e.g. Daleu et al. 2020).

Synoptic flow is also convection: latitudinally constrained by angular momentum, but still vertical and gravitational. Perhaps this same philosophical project can be brought to those scales? Time is still expected to naturally select ‘fit’ or efficient or (tautologically) winning configurations. Might the frequency of high-impact synoptic flow configurations like stationary waves or blocking be understood usefully in evolutionary terms, based on feedbacks in causal networks like Figs. 1-2 but with rotation and sphericity effects?

Darwinian evolution helped the descriptive biology of form turn its corner from *taxonomy* to *systematics*, greatly aided by genetics as a final arbiter. Could cloud pattern description be ripe for a similar leap? The holy grail would be if a fitness inferred from the frequency of occurrence of improbable patterns could be cross-checked by a measured or computed energy efficiency in simulations, then applies in observations. With these complementary ways of defining and measuring evolution, and a driving reason to care about how it plays out, that intellectual project in the tractable lifelessness of physics might even have something to offer back to the mighty, less tidy biological sciences. Being well-read in both game theory and fluid mechanics, the language model Sonnet 3.5 by Claude.ai offers a roadmap and mathematical framework for a research agenda in this area, an artifact published at https://bit.ly/3O5mKNx.

# Open science resources

All data used here are available public and online, including the Cloud Botany set (https://howto.eurec4a.eu/botany_dales.html) and NOAA Gridsat (DOI) infrared imagery. The MCS dataset (courtesy of Chen et al. 2023) was only lightly used to help select days in the rainy season. Codes and documentation are in Jupyter notebooks at https://github.com/brianmapes/EvolutionaryConvection, for instance the one with animations of Fig. 10 may be viewed using nbviewer at this shortened link: https://bit.ly/EvolvingConvectionSimulations. A collection of larger Cloud Botany derived products from this project, including a 103-page set of displays like Fig. 3 for all the simulations, is at https://bit.ly/CloudBotanyMapesDerived .

# Acknowledgements

Gratitude is expressed to reviewers of a sprawling first draft on ArXiv, notably Drs. Fiaz Ahmed and Jun-Ichi Yano, and long-striving thinkers on the complexity fringes of various science disciplines. Providers of excellent free datasets and software are too numerous to thank in these remarkable times. This material is based upon work supported by the National Science Foundation under Grant Numbers 2141492 and 2318221.